\documentclass[draftclsnofoot,onecolumn, 12pt]{IEEEtran}
\usepackage{amssymb}
\usepackage{setspace}
\usepackage{amsthm}
\usepackage{amsmath}
\usepackage{color}

\newtheorem{mydef}{Definition}
\newtheorem{prop}{Proposition}
\newtheorem{lemma}{Lemma}
\newtheorem{remark}{Remark}
\newcommand{\oo}[1]{o\left({#1}\right)}
\newcommand{\ooo}[1]{o({#1})}
\newcommand{\OO}[1]{O\left({#1}\right)}
\newcommand{\OOO}[1]{O({#1})}
\newcommand{\w}[1]{\omega({#1})}
\newcommand{\T}[1]{\Theta\left({#1}\right)}
\newcommand{\p}[1]{\mathrm{Pr}\left({#1}\right)}
\newcommand{\pp}[1]{\mathrm{Pr}({#1})}
\newcommand{\ee}[2][]{\mathrm{E}_{#1}\left({#2}\right)}
\newcommand{\eee}[1]{\mathrm{E}({#1})}

\newcommand{\I}[1]{\mathbf{1}_{#1}}
\newcommand{\D}[1]{\mathrm{d} {#1}}
\usepackage{color}

\ifCLASSINFOpdf
  \usepackage[pdftex]{graphicx}
\else
  \usepackage[dvips]{graphicx}
\fi

\begin{document}
\title{Large Overlaid Cognitive Radio Networks: From Throughput Scaling to Asymptotic Multiplexing Gain}

\author{Armin Banaei, Costas N. Georghiades, and Shuguang Cui% <-this % stops a space
\thanks{A. Banaei, C. Georghiades, and S. Cui are with the Department
of Electrical and Computer Engineering, Texas A\&M University.}}% <-this % stops a space

\maketitle

\begin{abstract}
%\boldmath
  We study the asymptotic performance of two multi-hop overlaid ad-hoc networks that utilize the same temporal, spectral, and spatial resources based on random access schemes. The primary network consists of Poisson distributed legacy users with density $\lambda^{(p)}$ and the secondary network consists of Poisson distributed cognitive radio users with density $\lambda^{(s)} = (\lambda^{(p)})^{\beta}$ ($\beta > 0$, $ \beta \neq 1$) that utilize the spectrum opportunistically. Both networks are decentralized and employ ALOHA medium access protocols where the secondary nodes are additionally equipped with range-limited \emph{perfect} spectrum sensors to monitor and protect primary transmissions. We study the problem in two distinct regimes, namely $\beta > 1$ and $0 < \beta < 1$. We show that in both cases, the two networks can achieve their corresponding stand-alone throughput scaling even without secondary spectrum sensing (i.e., the sensing range set to zero); this implies the need for a more comprehensive performance metric than just throughput scaling to evaluate the influence of the overlaid interactions. We thus introduce a new criterion, termed the \emph{asymptotic multiplexing gain}, which captures the effect of inter-network interferences with different spectrum sensing setups. With this metric, we clearly demonstrate that spectrum sensing can substantially improve primary network performance when $\beta > 1$. On the contrary, spectrum sensing turns out to be unnecessary when $\beta < 1$ and setting the secondary network's ALOHA parameter appropriately can substantially improve primary network performance.
\end{abstract}


% \fi
\IEEEpeerreviewmaketitle

\section{Introduction}
\label{sec: introduction}

\IEEEPARstart{G}{upta} and Kumar \cite{GuptaKumar} introduced a random network model for studying the throughput of large-scale static wireless networks where the network consists of $\lambda$ nodes that are independently and uniformly distributed over a unit-area disk. % to which it wants to communicate at $W$ bits per second, provided that the interference is sufficiently low.
Each node in the network could act as a source, a relay, or a destination, and each source node has a random destination in the network. The nodes have a common transmission range and each transmits to its one-hop neighbors in the direction of certain destination nodes.  They showed that a centralized time-slotted multi-hop transmission scheme can achieve a sum throughput scaling of $\T{\sqrt{\lambda/\log(\lambda)}}$\footnote{$f(\lambda) = \oo{g(\lambda)}$ means that $\lim f(\lambda)/g(\lambda) \rightarrow 0$ as $\lambda \rightarrow \infty$, $f(\lambda) = \OO{g(\lambda)}$ means that there exist positive constants $c$ and $M$ such that $f(\lambda)/g(\lambda) \leq c$ whenever $\lambda \geq M$, $f(\lambda) = \w{g(\lambda)}$ means that $\lim f(\lambda)/g(\lambda) \rightarrow \infty$ as $\lambda \rightarrow \infty$, $f(\lambda) = \T{g(\lambda)}$ means that both $f(\lambda) = \OO{g(\lambda)}$ and $g(\lambda) = \OO{f(\lambda)}$, $f(\lambda) \sim g(\lambda)$ means that $\lim f(\lambda)/g(\lambda) \to 1$ as $\lambda\to\infty$.}. Following \cite{GuptaKumar}, there has been a vast literature, e.g., \cite{closingthegap}-\cite{one-dimensionalmobility}, studying the asymptotic performance of single large-scale networks, all based on traditional static spectrum allocation schemes.

Conventional wireless communication systems will not be able to cope with the increasing demand for frequency spectrum in the future. Fortunately, although most of the usable frequency spectrum has already been allocated, they are scarcely utilized in different locations and at different times \cite{measurements}. In the seminal work of \cite{Mitola}, Mitola proposed \emph{cognitive radio} as a promising solution to utilize frequency spectrum more efficiently. The underlying idea is to let unlicensed users (secondary users) make use of the available temporal, spectral, or spatial opportunities over the licensed bands, while protecting the licensed users (primary users) by limiting the interference caused by the secondary users. Therefore, acute secondary interference management schemes are required by secondary users to maintain certain quality of service (QoS) for the primary network and achieve a reasonable performance for the secondary network in overlaid cognitive networks.

In this paper we study the asymptotic performance of multi-hop overlaid networks in which a primary ad-hoc network and a cognitive secondary ad-hoc network coexist over the same spatial, temporal, and spectral dimensions. In order to limit the secondary interference to the primary network, we adopt the \emph{dynamic spectrum access} \cite{DSAsurvey} approach, where secondary users opportunistically explore the white spaces detected using spectrum sensors. In \cite{vu}, Vu \emph{et al.} considered the throughput scaling law for single-hop overlaid cognitive radio networks, where a linear scaling law is  obtained  for the secondary network with an outage constraint considered for the primary network. In \cite{joen}, Jeon \emph{et al.} considered a  multi-hop cognitive network coexisting with  a primary network and assumed that the secondary nodes know the locations of all primary nodes (both primary transmitters and receivers). They showed that by defining a preservation region around each primary node and following time-slotted deterministic transmission protocols, both networks can achieve the same throughput scaling law as a stand-alone wireless network, while a vanishing fraction of the secondary nodes may suffer from a finite outage probability (as the number of the nodes tends to infinity). In \cite{yin}, the authors studied the throughput scaling and throughput-delay tradeoff with the same system model as in \cite{joen}, except that the secondary users only know the locations of the primary transmitters. By establishing preservation regions around primary transmitters, they showed that both networks could achieve the throughput scaling derived by Gupta and Kumar in \cite{GuptaKumar} without outage.

In all the previously mentioned papers, centralized deterministic schemes are used to achieve the feasible rates for both primary and secondary networks. Moreover, results are provided only when the secondary nodes are more densely distributed than the primary nodes. On the other hand, the desired autonomous feature of large wireless systems makes the use of a central authority to coordinate the primary/secondary users less appealing. In addition, in many practical situations, as the secondary users are opportunistic (or sporadic) spectrum utilizers, it is more likely that the secondary nodes are less densely distributed. In the literature, the asymptotic performance of traditional single-tier networks with distributed random access schemes has been studied, e.g., \cite{magicnumber}--\cite{Baccelli}. In \cite{magicnumber}, the performance of the slotted ALOHA protocol in a multi-hop environment was studied and the optimum transmission radius is derived to maximize the throughput for a random planar network. The spatial capacity of a slotted multi-hop network with capture was studied in \cite{aloha1}. In \cite{transport}, Weber \emph{et al.} derived the transmission capacity of wireless ad-hoc networks, where the transmission capacity is defined as the product between the maximum density of successful transmissions and their data rate, given an outage constraint. Baccelli \emph{et al.} \cite{Baccelli} proposed an ALOHA-based protocol for multi-hop wireless networks in which nodes are randomly located in an infinite plane according to a Poisson point process and are mobile according to a waypoint mobility model. They derived the optimum multiple access probability that achieves the maximum \emph{mean density of progress}.

In this work we consider decentralized ALOHA-based scheduling schemes for both primary and secondary networks in an overlaid scenario, where secondary users can only make use of localized information obtained via spectrum sensing to control their actions and limit their interferences to primary users. The distributed nature of ad-hoc networks and the passive property of primary receivers lead to uncertainties about the primary system state even with perfect spectrum sensing. As such, we focus on the case where the secondary users are able to \emph{perfectly detect} the primary user signals when the primary transmitters are within a certain range. In particular, we study the asymptotic performance of the two overlaid networks, where we start with the throughput scaling laws, and then introduce a new metric called \emph{asymptotic multiplexing gain} that further quantifies the performance tradeoff between the two networks. We do so under two scenarios, i.e., the secondary network is denser vs. sparser than the primary network, and identify their key differences. To the best of our knowledge, this is the first time that the achievable rates for overlaid cognitive networks with random access schemes is studied, where the secondary network could be either denser or sparser than the primary network.

The rest of the paper is organized as follows. Section \ref{sec:systemmodel} introduces the mathematical model, notations, and definitions. In Section \ref{sec:singletier} we consider the \emph{spatial throughput} of the single-tier network. Section \ref{sec:CognitiveNetworkThroughputScaling} studies the cognitive overlaid scenario and addresses the tradeoff between the primary and secondary networks by introducing the notion of asymptotic multiplexing gain (AMG). In particular, we show that both networks can achieve their corresponding single-tier throughput scaling regardless of the setting for the spectrum sensing range. However, for the case with a denser secondary network, spectrum sensing can improve the primary network AMG; whereas, for the case with a sparser secondary network, the spectrum sensors turn out to be redundant and the primary network AMG can be enhanced by reducing the medium access probability of secondary users while maintaining a non-trivial sum-throughput for the secondary network. Section \ref{sec:conclusion} concludes the paper.

%\hfill January 11, 2007

\section{System Model and Definitions}
\label{sec:systemmodel}

Consider a circular area $A$ in which a network of primary nodes and a network of secondary nodes share the same temporal, spectral, and spatial resources\footnote{All the results will carry over to any smooth and convex region with some minor considerations.}. Both primary and secondary nodes are distributed according to Poisson point processes with densities $\lambda^{(p)}$ and $\lambda^{(s)}=(\lambda^{(p)})^{\beta}$ ($\beta > 0, \beta \neq 1$), respectively. Let $\phi^{(p)} = \{X_i^{(p)}\}$ and $\phi^{(s)} = \{X_i^{(s)}\}$ denote the (Cartesian) coordinates of a realization of the primary and secondary nodes. As mentioned earlier, the primary users are legacy users, and thus have a higher priority to access the spectrum; the secondary users can access the spectrum opportunistically (based on the spectrum sensing outcome) as long as they abide by ``certain'' interference constraints.

Throughout this paper we denote the parameters associated with the primary and the secondary users with superscripts $(p)$ and $(s)$, respectively; e.g., $R_I^{(p)}$ denotes the interference range from a primary transmitter to a primary receiver and $R_I^{(s)}$ denotes the interference range from a secondary transmitter to a secondary receiver.

Each primary receiver tries to decode the signal from its intended transmitter located within $R_r^{(p)}$ radius and is prone to interference from other primary and secondary transmitters within $R_I^{(p)}$ and $R_I^{(sp)}$ radii, respectively. Likewise, a secondary receiver tries to decode the signal from its intended transmitter located within $R_r^{(s)}$ radius and is prone to interference from other secondary and primary transmitters within $R_I^{(s)}$ and $R_I^{(ps)}$ radii, respectively. Furthermore, due to certain cognitive features\footnote{e.g., acquiring knowledge about primary messages and utilizing joint encoding techniques to partially mitigate primary interference.}, we may assume that the cognitive secondary receivers are more robust against primary interferences than primary receivers, i.e., $R_I^{(ps)} \leq R_I^{(p)}$ (with $R_I^{(ps)} = \OOO{R_I^{(p)}}$); also, primary receivers are more sensitive to the secondary interference than secondary receivers\footnote{e.g., due to possible multiuser cooperation among secondary users.}, i.e., $R_I^{(sp)} \geq R_I^{(s)}$ (with $R_I^{(sp)} = \OOO{R_I^{(s)}}$). In addition, it is reasonable to assume that the interference range is no less than the transmission range for both networks, i.e., $R_I^{(p)} = \sqrt{1+l^{(p)}}R_r^{(p)}$ and $R_I^{(s)} = \sqrt{1+l^{(s)}}R_r^{(s)}$ for some constants $l^{(p)},l^{(s)} \geq 0$. Further, secondary nodes are equipped with perfect spectrum sensors that can reliably detect the primary user signals (i.e., the existence of transmitting primary users) within $R_D$ radius. %The relationship among all aforementioned ranges is encapsulated in Fig. \ref{fig:Radii}.

%\begin{figure}
%  \centering
%  \includegraphics[width=4 in]{Radii.eps}
%  \caption{Primary and secondary network parameters, where $t_x^{(p)}$ and $r_x^{(p)}$ denote a primary transmitter and receiver pair, and $t_x^{(s)}$ and $r_x^{(s)}$ denote a secondary transmitter and receiver pair, respectively.}
%  \label{fig:Radii}
%\end{figure}

Let $|A|$ denote the area of region $A$ and $B_{R}(\cdot)$ denote a full disk with radius $R$ centered at $(\cdot)$, which could be either the polar coordinates in the form of $(r,\varphi)$ or the location of a node $X$ in the form of $(X)$. We interpret $B_{R_1}(r_1,\varphi_1)-B_{R_2}(r_2,\varphi_2)$ as the remaining region of a disk with radius $R_1$ centered at polar coordinates $(r_1,\varphi_1)$ excluding the overlapping region with another disk with radius $R_2$ centered at $(r_2,\varphi_2)$. Furthermore, given measurable sets (or events) $\sigma_1$ and $\sigma_2$ we denote by $\overline{\sigma}_1$ the complement of event $\sigma_1$ and denote by $\sigma_1\sigma_2 := \sigma_1\cap \sigma_2$ their intersection.

For the transmission protocols in both networks, the time axis is slotted and the slot duration is defined as the time required to transmit a packet in the system, where all packets are assumed to be of the same size. In the following, we outline the primary and secondary network protocols, both based on the slotted ALOHA structure.

\subsection{Primary Network Protocol}
\label{subsec:PrimaryNetworkProtocol}

Each primary node picks a destination uniformly at random among all other nodes in the primary network. Communication occurs between a primary source-destination (S-D) pair through a single-hop transmission if they are close enough, or through multi-hop transmissions over intermediate relaying nodes if they are far apart. In this manner, each primary node might act as a source, destination or a relay, and always has a packet to transmit (which is either its own packet or a packet being relayed). We assume that each node has an infinite queue for packets where the first packet in the queue is transmitted with probability $q^{(p)}$ (the ALOHA parameter). The selection of relaying nodes along the (multi-hop) routing path is governed by a variant of geometric routing schemes, \cite{MFR}--\cite{analys}, namely \emph{the random $\frac{1}{2}$disk routing scheme}\footnote{We choose the random $\frac{1}{2}$disk routing scheme mainly for tractability and simplicity in mathematical characterization. However, the solution techniques developed in this paper can be used (with some modifications) to study other variants of geographical routing schemes, such as MFR, NFP, DIR, etc.} as discussed in Section \ref{subsec:routingpath}.

\subsection{Secondary Network Protocol}
\label{subsec:SecondaryNetworkProtocol}

Similar to the primary network, each secondary node picks a destination uniformly at random among all other nodes in the secondary network. Each secondary node has an infinite queue for packets with the first one in the queue transmitted with probability $q^{(s)}$, \emph{whenever} the channel is deemed idle: In particular, each secondary user senses the channel for primary activities prior to a transmission initiation and commences the transmission of the first packet in the queue with probability $q^{(s)}$ whenever there are no primary transmitters detected within $R_D$ radius. Setting $R_D = 0$ implies that secondary nodes always initiate transmissions with probability $q^{(s)}$ regardless of the primary channel occupancy status. The secondary network utilizes a similar routing scheme to that in the primary network.

\subsection{Random $\frac{1}{2}$Disk Routing Scheme}
\label{subsec:routingpath}

Since both primary and secondary networks utilize the same routing scheme, in this section we introduce our routing scheme for a generic wireless ad-hoc network (omitting the superscripts $(p)$ and $(s)$). Throughout the paper, we assume that both primary and secondary networks possess the following property: each network node has at least one relaying node in every direction with high probability; this is a sufficient condition for the existence of routing paths (with finite lengths) between any arbitrary source-destination pair in the network and can be guaranteed asymptotically almost surely if $R_r= K \sqrt{\log\lambda/\lambda}$ for a large enough constant $K$ (c.f. \cite{halfdisk}, Theorem 1).

Consider an arbitrary packet $b$ for a source-destination pair that is $h$-distance apart. We set the destination node at the origin and assume that the routing path starts from the source node at $X_0 = (-h,0)$, where $X_n$ is the (Cartesian) coordinate of the $n^{\textrm{th}}$ relay node along the routing path and $r_n:=\|X_n\|$ is the (Euclidean) distance of the $n^{\textrm{th}}$ relay node from the destination.

\begin{figure}
  \centering
  \includegraphics[width=4 in]{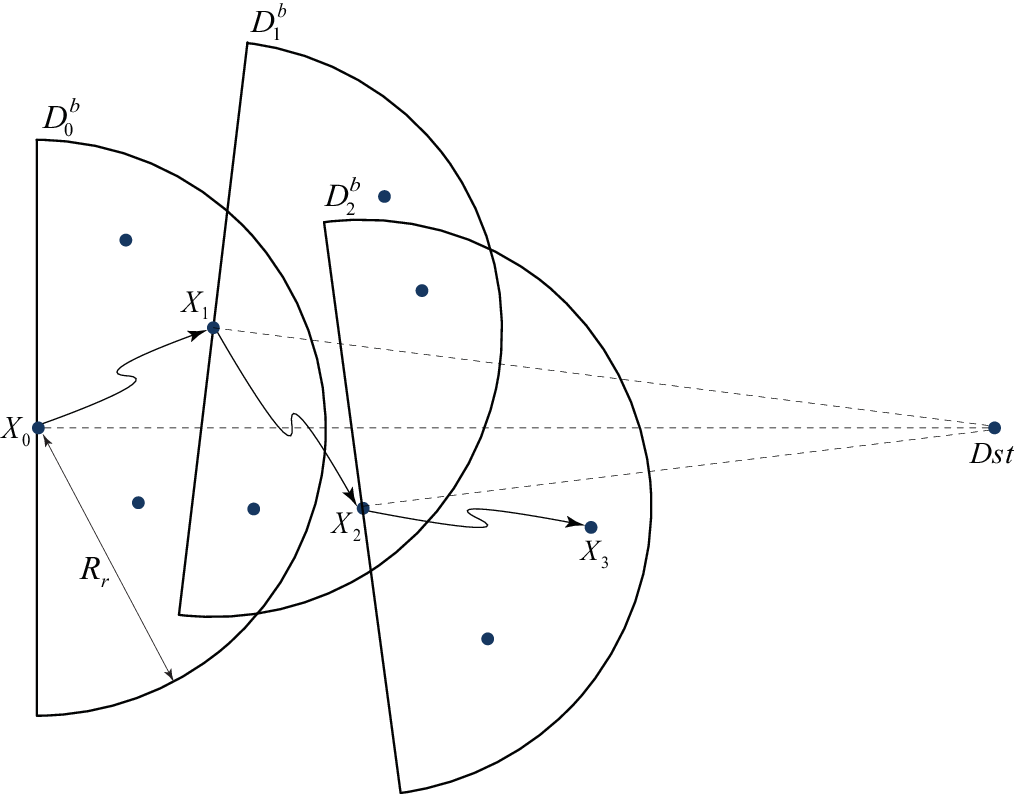}
  \caption{Evolution of the random $\frac{1}{2}$disk routing path.}
  \label{fig:routingevolution}
\end{figure}
More specifically, the routing path starts at the source node $X_0 = (-h,0)$ with its transmission $\frac{1}{2}$disk $D_0^b$ that is a $\frac{1}{2}$disk with radius $R_r$ centered at $X_0$ and oriented towards the destination at $(0,0)$. The next relay $X_1$ is selected at random from nodes contained in $D_0^b$. This induces a new $\frac{1}{2}$disk $D_1^b$, centered at $X_1$ and oriented towards the destination. Relay $X_2$ is selected randomly among the nodes in $D_1^b$, and the process continues in the same manner until the destination is within the transmission range. We claim that the routing path converges (or is established) whenever it enters the transmission/reception range of the final destination, i.e., $r_{\tau} \leq R_r$, for some $\tau \in \{0, 1, 2, \ldots\}$. In Fig. \ref{fig:routingevolution}, we illustrate the progress of a packet towards its destination. We define the \emph{progress} at the $n^{\textrm{th}}$ hop of the routing path as $Y_n := \|X_n\| - \|X_{n+1}\| = r_n - r_{n+1}$.

\subsection{Spatial Throughput}
\label{subsec:SpatialThroughput}

In this paper we adopt a notion of throughput similar to \emph{mean spatial density of progress} in \cite{Baccelli}.

\begin{mydef}
\label{def:spatialthroughput}
We define the \textsl{spatial throughput} of the network as the mean total progress of all successfully transmitted packets in the whole network over a single hop. More specifically, let $b$ be the packet at the head of queue of node $X\in\phi$, $Y_{X}^b$ be the progress of packet $b$ at node $X$, and $\Lambda_{X}^b$ be the event of successful transmission of packet $b$ at node $X$. Then the spatial throughput of the network is defined as\footnote{In this paper we ignore the edge effects, i.e., we assume that the location of network nodes in $B_{R}(X)$ is uniformly distributed irrespective of the location of $X$. Essentially, we are ignoring the fact that the portion of disks around edge nodes that fall outside of the network region do not contain any other nodes.}
\begin{equation}
\label{eq:def1_spatialthroughput}
C := \lambda |A| \ee{Y_{X}^b \I{\Lambda_{X}^b}}\,,
\end{equation}
where $\I{}$ is the indicator function and $\eee{}$ is the expectation operator taken over all realizations of the network nodes, source-destination pair assignments, and the routing paths between S-D pairs.
\end{mydef}

\begin{figure}
  \centering
  \includegraphics[width=4 in]{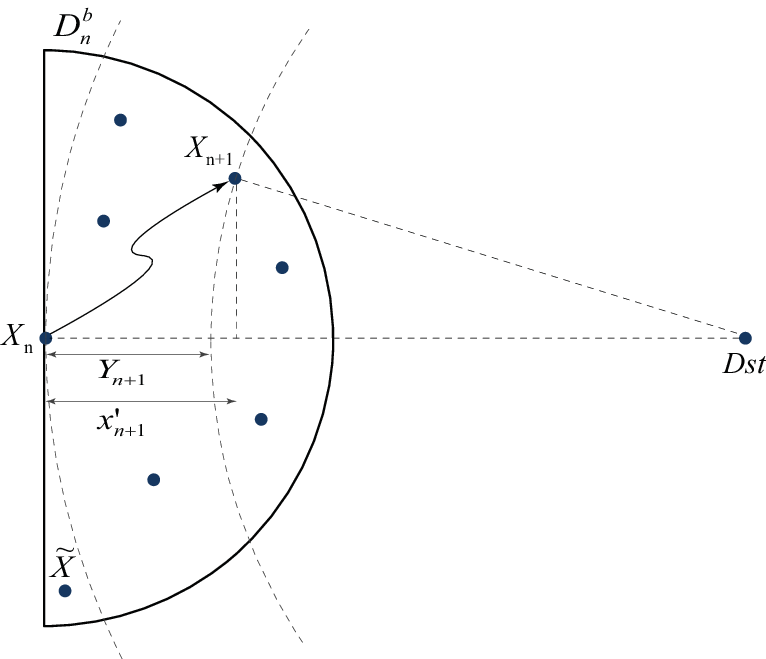}
  \caption{Progress of the packet at the $n^{\textrm{th}}$ hop. $Y_{n+1}$ is the decrement in the radial distance of a packet to its destination and $x'_{n+1}$ is the decrement in the distance of the projection of the packets position on the line connecting the transmitting node and the destination.}
  \label{fig:progress}
\end{figure}

There are two key differences between our notion of throughput and the mean spatial density of progress. The first difference lies in the fact that in the mean spatial density of progress only a typical snapshot of the network is considered and the progress is computed only for the typical realization of the local neighborhood of a transmitting node. However, in our notion of throughput we consider the whole routing path of a packet and compute the mean progress of the packet over a single hop along that path. In other words, we are computing the expected progress of packets over both time and space. The second difference between our notion of throughput and the mean spatial density of progress stems from the definition of the progress, where in \cite{Baccelli} the progress is defined to be the decrement in the distance of the packet's position projected on the line connecting the transmitting node and the destination, whereas in this paper we define the progress to be the decrement in the radial distance of a packet to its destination, as shown in Fig. \ref{fig:progress}. In order to highlight the difference between these two definitions, consider the following exaggerated example.

Assume a (very unfortunate) realization of the routing path where at each hop a node in the upper/lower corner of the transmission $\frac{1}{2}$disk is chosen as the next relay (e.g., $\tilde{X}$ in Fig. \ref{fig:progress}). Over this path, the packet gets farther away from the destination at each hop and should never reach the destination; this is an intuitive result that our definition of progress complies with. However, according to the projected distance progress definition in \cite{Baccelli}, at each hop, the packet has made a positive drift towards the destination and should eventually reach the destination. Furthermore, based on the projected distance progress, the progress of a packet towards its destination is i.i.d. over all relay nodes. This means that the packet progress is independent of the distance from the transmitting node to the destination. However, as we show later, the packet distance from the destination decreases more (on average), when it is farther away from the destination, and decreases less as the packet gets closer to the destination (c.f. \eqref{eq:gbound}). This suggests that the packet progress is a function of its relative position to its destination and the current distance from the packet to the destination should be considered in evaluating the progress at each hop. In the next section, we determine the spatial throughput for the stand-alone primary and secondary networks and provide some interpretations for this metric.

\section{Single Network Throughput Scalings}
\label{sec:singletier}

In this section we consider the spatial throughput of a single-tier network when no other networks are overlaid. This serves as a performance benchmark for the overlaid case discussed in the next section. The following lemma provides us with an equivalent definition and a method of computing the spatial throughput for our system.
\begin{lemma}[Separation Principle]
\label{lemma:separation_principle}
Consider the single-tier version of the wireless ad-hoc network defined in Section \ref{sec:systemmodel}. The spatial throughput of  such a network equals the product between the expected number of simultaneously successful transmissions in the whole network and the average progress of a typical packet over a single-hop transmission. Specifically, the spatial throughput of the network can be obtained as
\begin{equation}
\label{eq:def2_spatialthroughput}
C = \lambda |A| \ee{Y^b_X}\p{\Lambda^b_X}\,,
\end{equation}
where $\Lambda_X$ and $Y^b_X$ are defined in Definition \ref{def:spatialthroughput} and the expectation is taken over all realizations of the network nodes, S-D assignments, and the routing paths between S-D pairs. %$\pp{}$ is the probability measure and
\end{lemma}
\begin{proof}
Let $b$ be the packet at the head of node $X$'s queue at an arbitrary time slot. Note that $b$ and $Y^b_X$ are random variables dependent on the specific realization of the network nodes, the S-D assignments, and the routing path establishment with the random $\frac{1}{2}$disk routing scheme. Assume that $X_0$ and $X_{\nu^b+1}$ are the source and destination of packet $b$ respectively, where $\nu^b+1$ is total number of hops that $b$ traverses over. Let $\{X_1, X_2, \ldots X_{\nu^b}\}$ be the nodes that $b$ hops over. We have
\begin{align*}
\ee[\nu^b]{Y^b_X\I{\Lambda^b_X}} &= \ee{Y^b_X\I{\Lambda^b_X}\I{\{X=X_n: n = 0, \ldots, \nu^b\}}} \\
&= \frac{1}{\nu^b+1}\sum_{n=0}^{\nu^b}\ee{Y^b_{X_n}\I{\Lambda^b_{X_n}}}\,,
\end{align*}
where we define $\ee[X]{Y} := \ee{Y\mid X}$. Therefore, we can reformulate \eqref{eq:def1_spatialthroughput} as
\begin{equation}
\label{eq:betterdef}
C = \lambda |A| \ee{\ee[\nu^b]{Y^b_X\I{\Lambda^b_X}}} = \lambda |A| \ee{\frac{1}{\nu^b+1}\sum_{n=0}^{\nu^b}\ee{Y^b_{X_n}\I{\Lambda^b_{X_n}}}}\,.
\end{equation}

Now, consider the transmission of packet $b$ from node $X_n$ to $X_{n+1}$. Packet $b$ is successfully transmitted/relayed if:
\begin{enumerate}
  \item[I)] Node $X_n$ initiates a transmission according to the ALOHA protocol with probability $q$ (denoted by event $\Lambda^b_{1,X_n}$).
  \item[II)] For any node $X_{n+1}$ that is selected as the next relay for $b$ according to the random $\frac{1}{2}$disk routing scheme, we have that neither $X_{n+1}$ nor any other nodes contained in its interference range $B_{R_I}(X_{n+1})$, except for $X_n$, initiate a transmission (denoted by event $\Lambda^b_{2,X_{n+1}}$).
\end{enumerate}

Note that since we assumed $R_I \geq R_r$, $\Lambda^b_{2,X_{n+1}}$ also implies that in the event of successful transmission no two nodes transmit packets to $X_{n+1}$ at the same time. Moreover, $\Lambda^b_{X_n}$ only depends on the multiple access decisions of $X_n$, $X_{n+1}$, and the nodes that are contained in the interference range of $X_{n+1}$. All these nodes initiate transmissions independent of each other and independent of all previous transmission attempts. Together with the fact that all network nodes always have a packet to transmit, we conclude that $\pp{\Lambda^b_{X_n}}$ only depends on the number of nodes contained in the interference range of the next relay node. Hence,  due to the homogeneity of the underlying Poisson point process of the network nodes, $\pp{\Lambda^b_{X_n}}$ is only a function of the area of $B_{R_I}(X_{n+1})$, and is independent of the realization of $X_{n+1}$. In other words, $\{\Lambda^b_{X_n}\}_{b,n}$ are identically distributed (but possibly correlated) collection of random variables, and are independent of $X_n$, $X_{n+1}$, and consequently $Y^b_{X_n}$. From \eqref{eq:betterdef} we get
\begin{align*}
C &= \lambda |A| \ee{\frac{1}{\nu^b+1}\sum_{n=0}^{\nu^b}\ee{Y^b_{X_n}}}\p{\Lambda^b_{X_n}}\\
&= \lambda |A| \ee{Y^b_{X}}\p{\Lambda^b_{X}}
\end{align*}
\end{proof}

As a consequence of Lemma \ref{lemma:separation_principle}, we could derive the spatial throughput of the network by separately determining the probability of a successful one-hop transmission and the average progress for a typical packet $b$ at a typical node $X$. Based on the proof of Lemma \ref{lemma:separation_principle} we have
\begin{align}
\label{eq:successprob}
\p{\Lambda^b_{X}} &= \ee{\sum_{X_j\in D^b_{X}}\frac{q(1-q)^{n_{X_j}+1}\I{n_{X_j}>0}}{n'_X}} \nonumber \\
&= q(1-q)e^{-\lambda q |B_{R_I}|} \left(1-e^{-\lambda (1-q) |B_{R_I}|}\right) \nonumber\\
&= q(1-q)e^{-\lambda q \pi R_I^2} \left(1-e^{-\lambda (1-q) \pi R_I^2}\right)\,,
\end{align}
where $n'_X \sim\mathrm{Pois}(\lambda |D^b_X|)$ is the number of nodes in $D^b_X$ and $n_{X_j} \sim \ \mathrm{Pois}(\lambda |B_{R_I}|)$ is the number of nodes in the interference range of $X_j$ (excluding $X_j$ and $X$).

\begin{figure}
  \centering
  \includegraphics[width=4 in]{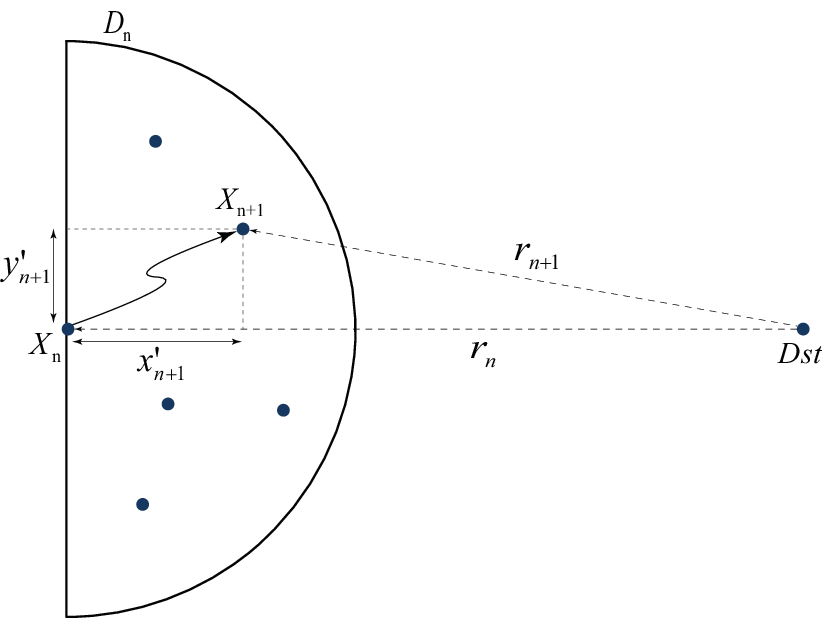}
  \caption{Distance between the next relay and the current node projected onto to the local coordinates at the current node.}
  \label{fig:routingschemebound}
\end{figure}

In order to derive the average packet progress we need some more nomenclature and intermediate results. Consider a packet $b$. To simplify the notation we drop the superscripts associated with this packet. According to \cite{halfdisk} (c.f., Proposition 1), we can (approximately) model the distance $\{r_n\}$ of packet $b$ to its destination as a Markov process solely characterized by its progress $\{Y_n\}$. Let $\{X_n\}$ be the set of nodes that $b$ hops over, and let $(x'_{n+1},y'_{n+1})$ be the projection of $X_{n+1}-X_n$ onto the \emph{local} Cartesian coordinates with node $X_n$ as the origin and the $x$-axis pointing from $X_n$ to the destination node as shown in Fig. \ref{fig:routingschemebound}. Hence, we have
\begin{equation}
\label{eq:alternativeformulation}
r_{n+1} = \sqrt{(r_n-x'_{n+1})^2 + y'^{2}_{n+1}}\,.
\end{equation}

According to \cite{halfdisk} (Proposition 1), $X_{n+1}$ is uniformly distributed on $D_n$ for a large enough $\lambda$; hence $\{(x'_{n},y'_{n})\}$ is an i.i.d. sequence of random variables with ranges $x'_{n} \in [0,R_r]$ and $y'_{n} \in [-R_r, R_r]$ for all $n$, whenever $\lambda$ is large enough.

Define $\nu^{(h)}_{r} := \inf\{n:r_n \leq r,\, r_0 = h\}$, $R_r \leq r\leq h$, to be the index of the first relay node closer than $r$ to the destination when the source and destination nodes are $h$-distance apart. Hence, $\nu^{(h)}_{R_r}+1$ represents the length (or hop-count) of the routing path. In \cite{halfdisk} we prove that under certain conditions for $R_r$, $\nu^{(h)}_{R_r}$ is finite asymptotically almost surely. Note that $\nu^{(h)}_{r}$ is a stopping time \cite{Resnik} and
\begin{equation*}
r-R_r \le r_{\nu^{(h)}_r}\le r\,.
\end{equation*}

Furthermore, let $g(r,x',y'):=\sqrt{(r-x')^2+y'^2}-r$. Observe that $g$ is a non-decreasing function over $r > R_r$, for fixed $(x',y')$, and $-g(r_n,x'_{n+1},y'_{n+1}) = Y_n$ (the progress at the $n^{\textrm{th}}$ relay). Thus, for $n<\nu^{(h)}_r$, we have $r_n > r$ and
\begin{align}
\label{eq:gbound}
-x'_{n+1} \le r_{n+1}-r_n &= g(r_n,x'_{n+1},y'_{n+1}) \nonumber\\
&\le g(r,x'_{n+1},y'_{n+1}) \le -x'_{n+1}+\frac{R_r^2}{2r}\,.
\end{align}

Hence, for a source-destination pair that is $h$-distance apart ($r_0 = h$), we have
\begin{subequations}
\begin{align}
r-R_r &\le r_{\nu^{(h)}_r}\le h+\sum_{n=0}^{\nu^{(h)}_r}g(r,x'_{n+1},y'_{n+1})\,, \label{eq:ba} \\
h+\sum_{n=0}^{\nu^{(h)}_r}(-x'_{n+1}) &\le r_{\nu^{(h)}_r}\le r\,.  \label{eq:bb}
\end{align}
\end{subequations}
which together with \eqref{eq:gbound} yields
\begin{align*}
\label{eq:progressbound}
\ee{\frac{h-r}{\nu^{(h)}_r+1}-\frac{R_r^2}{2r}} \leq \ee{Y^b_X} &= \ee{\frac{1}{\nu^{(h)}_r+1}\sum_{n=0}^{\nu^{(h)}_r} Y_n} \\
&= \ee{\frac{1}{\nu^{(h)}_r+1}\sum_{n=0}^{\nu^{(h)}_r} -g(r_n,x'_{n+1},y'_{n+1})} \leq \ee{\frac{h-r+R_r}{\nu^{(h)}_r+1}}\,,
\end{align*}
where the expectation is taken over all network, S-D assignment, and routing path realizations. Now let $S_m := \sum_{n=1}^{m}x'_n$ with $S_0 = 0$, and $\eta(z) := \eee{e^{z x'_n}}$. We know that $\exp(zS_m-m\log(\eta(z)))$ is a positive martingale, with value $1$ at $m = 0$ \cite{Resnik}. Hence, recalling \eqref{eq:bb}, we have
$$\ee{e^{z(h-r)-(\nu^{(h)}_r+1)\log(\eta(z))}} \leq \ee{e^{zS_{\nu^{(h)}_r+1}-(\nu^{(h)}_r+1)\log(\eta(z))}} \leq 1\,.$$

This implies
\begin{equation}
\ee{e^{-(\nu^{(h)}_r+1)\log(\eta(z))}} \leq e^{-z(h-r)}\,.
\end{equation}

Using Jensen's inequality and the monotone convergence theorem \cite{Resnik}, it is easy to show that
\begin{align}
\frac{1}{\ee{\nu^{(h)}_{R_r}+1\mid h}} \leq \ee{\frac{1}{\nu^{(h)}_{R_r}+1}\bigg| h} &= \int_0^{\infty}\ee{e^{-t(\nu^{(h)}_{R_r}+1)}}\D{t} \nonumber\\
&\leq \int_0^{\infty} e^{-z(h-r)}\D{(\log(\eta(z)))} \nonumber\\
%&= \int_0^{\infty} \frac{\eta'(u)}{\eta(u)}e^{-u(h-r)}\D{u} \nonumber\\
&= \int_0^{\infty} \frac{\ee{x'_n e^{zx'_n}}}{e^{zx'_n}}e^{-z(h-r)}\D{z} \nonumber\\
&\leq \int_0^{\infty} \ee{x'_n} e^{-z(h-r-R_r)}\D{z} \nonumber\\
&= \frac{\ee{x'_n}}{h-r-R_r}\,.
\end{align}

Finally, choosing $r = R_r(1+\sqrt{\frac{h}{R_r}})$, we can determine the average progress of a typical packet at a typical node $X$ by
\begin{equation}
\label{eq:progress}
\ee{Y^b_X} = \frac{4R_r}{3\pi} + \OO{R_r^{3/2}} \sim  \frac{4R_r}{3\pi}\,,
\end{equation}
where we have used the facts that $\eee{\nu^{(h)}_{R_r}\mid h} \sim \frac{h}{\ee{x'_n}}$ and $\ee{x'_n} = \frac{4R_r}{3\pi}$ (c.f. \cite{halfdisk}). Combining \eqref{eq:successprob} and \eqref{eq:progress} we obtain the spatial throughput of the single-tier network as
\begin{subequations}
\begin{align}
C &\sim \lambda |A| \ee{\frac{h}{\nu_{R_r}^{(h)}+1}} \label{eq:1tierthroughput_-1}\p{\Lambda^b_X}\\
&= \frac{4|A|}{3\pi} \lambda q(1-q)R_r e^{-\lambda q \pi R_I^2} \left(1-e^{-\lambda (1-q) \pi R_I^2}\right) \label{eq:1tierthroughput_0} \\
&= \T{\sqrt{\frac{\lambda}{\log(\lambda)}}}\,, \label{eq:1tierthroughput}
\end{align}
\end{subequations}
when $q = \OO{1/\log(\lambda)}$ and $R_r = \OO{\sqrt{\log(\lambda)/\lambda}}$. Observe that based on \eqref{eq:1tierthroughput_0}, one can show that $q = (\lambda \pi R_I^2)^{-1}$ maximizes the spatial throughput of the network (when $\lambda$ is large) and $q = \OO{1/\log(\lambda)}$ is a \emph{necessary} condition for $C$ to be asymptotically nontrivial.

\begin{remark}
Observe that if the network is \emph{stable}, the spatial throughput of the network equals the expected number of packet-meters that the network delivers to the destinations at each time slot, which is equivalent to the transport capacity defined in \cite{GuptaKumar}. The network is stable if the rate at which new packets are generated is equal to the rate at which packets are delivered to their respective destinations. In other words, the queue length of all network nodes is almost surely finite and packets are not being stored indefinitely in some nodes in the network. Intuitively, when the network is stable, there are $\lambda |A| \p{\Lambda_X}$ successful one-hop transmissions occurring in the whole network in each time slot; however, due to relaying, only $\eee{h/\nu_{R_r}^{(h)}+1}$ of these successful transmissions (on average) contribute to the throughput and the rest are only the retransmissions of relayed packets\footnote{The temporal analysis of the system is beyond the scope of this paper and will be discussed in a future work.}.
\end{remark}

We denote the spatial throughput of stand-alone primary and secondary networks by $C^{(p)}$ and $C^{(s)}$ respectively; i.e., $C^{(p)}$ (or $C^{(s)}$) equals the single-tier spatial throughput expression in \eqref{eq:1tierthroughput_0} with primary (or secondary) network parameters substituted. We will show in Section \ref{sec:CognitiveNetworkThroughputScaling} that even when we have two networks sharing the same resources and the secondary network accesses the spectrum without sensing (as if the primary tier is not present), both networks can still achieve the above throughput scaling. This suggests that throughput scaling alone is not adequate to evaluate the performance of large-scale overlaid networks, as it masks the effect of mutual interference between the two networks. Intuitively, when the secondary users try to access the spectrum more aggressively, the primary network throughput should degrade. However, it turns out that the augmented interference from secondary users only causes a constant penalty to the primary throughput in the asymptotic sense such that the scaling law by itself cannot reflect this effect.

To quantify the effect of mutual interference between the two networks, we define a new measure, \emph{asymptotic multiplexing gain} (which should be a function of the spectrum sensing range at the secondary nodes), to characterize the protection vs. competition tradeoff between the two networks.

\begin{mydef}
Assume that the throughput $C(\lambda)$ of a network scales as $\T{f(\lambda)}$; we define the \textsl{Asymptotic Multiplexing Gain (AMG)} of the network as the constant $\chi$ such that:
\begin{equation}
\chi := \lim_{\lambda\to\infty} \frac{C(\lambda)}{f(\lambda)}.
\end{equation}
\end{mydef}

Note that the exact value of $\chi$ may not be always computable, but its bounds always are. As such, we can define a \emph{partial ordering} \cite{bredin} on the set of all network throughputs. Specifically, consider two networks $N_1$ and $N_2$ with throughputs $C_{N_1}$ and $C_{N_2}$, and asymptotic multiplexing gains $x_1 \leq \chi_{N_1} \leq y_1$ and $x_2 \leq \chi_{N_2} \leq y_2$. We say $C_{N_1} \preceq C_{N_2}$ if and only if $C_{N_1}/C_{N_2} = \oo{1}$, or $y_1 \leq x_2$ when $C_{N_1}/C_{N_2} = \OO{1}$\footnote{This definition closely resembles Lexicographic ordering \cite{bredin}.}. From a different perspective, if we plot $C(n)$ over $f(n)$ for asymptotically large $n$, AMG is nothing but the slope of the throughout scaling curve, hence the connotation ``multiplexing gain''; and it is intuitive to always desire a large AMG.

Accordingly, we can determine the single-tier network AMG in the absence of the other network as:
\begin{equation}
\label{eq:singletierAMG}
\chi= \frac{4|A|e^{-1}}{3\pi^2\sqrt{1+l}}\,,
\end{equation}
when $q = (\lambda \pi R_I^2)^{-1}$ and $R_I/R_r = \sqrt{1+l}$.

\section{Overlaid Cognitive Network Spatial Throughput}
\label{sec:CognitiveNetworkThroughputScaling}

In this section we consider the case where both primary and secondary networks are present in the overlaid fashion under two distinct scenarios: one with the secondary network being denser than the primary network ($\beta > 1$) and the other with the primary network being denser ($\beta < 1$). As shown later, the impact of each tier on the spatial throughput of the other tier is materialized in the reduction of expected number of successful one-hop transmissions.

The distinctive feature of the overlaid cognitive network is that the secondary users are allowed to transmit only if they detect no primary transmitters within an $R_D$ radius. The possible overlap between the detection ranges of secondary users correlates their medium access decisions, which consequently, correlates the successes of one-hop transmissions with the Euclidean hop-lengths in both primary and secondary networks. Therefore, in the overlaid scenario, the separation principle (Lemma \ref{lemma:separation_principle}) is no longer directly applicable;  this makes the characterization of the primary and secondary network spatial throughputs challenging.

In the following two subsections we derive the spatial throughputs of the overlaid cognitive radio networks. The analysis closely follows that in the previous section, however, with proper modifications that take into account the opportunistic access mechanism adopted by secondary users and the extra inter-network interferences.

\subsection{Throughput Analysis for the Primary Network}
\label{subsec:PUspatialThroughput}

Let $\Lambda_{X^{(p)}_n}$ be the event of successful transmission for primary packet $b$ from a primary node $X^{(p)}_n$ to the next relay $X^{(p)}_{n+1}$ in the presence of the secondary network\footnote{Henceforth, we drop the superscript $b$ for brevity.}. We have that $\Lambda_{X^{(p)}_n}$ happens if events $\Lambda_{1,X^{(p)}_n}$, $\Lambda_{2,X^{(p)}_{n+1}}$, and $\Lambda_{3,X^{(p)}_{n+1}}$ all happen. As in the proof of Lemma \ref{lemma:separation_principle}, $\Lambda_{1,X^{(p)}_n}$ denotes the event that $X^{(p)}_n$ initiates a transmission, $\Lambda_{2,X^{(p)}_{n+1}}$ denotes the event that neither $X^{(p)}_{n+1}$ nor any primary nodes contained in $B_{R_I^{(p)}}(X^{(p)}_{n+1})$, except $X^{(p)}_n$, initiate a transmission, and $\Lambda_{3,X^{(p)}_{n+1}}$ denotes the event that there are no secondary transmitters within inter-network interference range $R_I^{(sp)}$ of $X^{(p)}_{n+1}$.

Recall that we require the secondary network to be transparent to the primary network. Hence, we assume that primary users utilize the same medium access probability as if the secondary tier was not present, i.e., we set $q^{(p)} = (\lambda^{(p)} \pi (R_I^{(p)})^2)^{-1}$. On the other hand, each secondary transmitter initiates transmission with probability $q^{(s)}$ only if it detects the channel as \emph{idle}, i.e., when there are no primary transmitters within $R_D$ radius. Therefore, if $X^{(p)}_n$ initiates a transmission, all secondary users in $B_{R_D}(X^{(p)}_n)$ would refrain from transmission. As such, to compute the probability of successful transmission for the primary network, we only need to consider the possible inter-network interference from the secondary nodes in $B_{R_I^{(sp)}}(X^{(p)}_{n+1})-B_{R_D}(X^{(p)}_n)$. From this we observe the following two facts:
\begin{itemize}
  \item[i)] The likelihood of a secondary user interfering with the transmission from $X^{(p)}_{n}$ to $X^{(p)}_{n+1}$ decreases as $R_D$ increases. Thus, the probability of successful transmission for a primary user is an increasing function of $R_D$. Setting $R_D = R_I^{(sp)} + R_r^{(p)}$ guarantees zero interference from the secondary network to the primary network since all the secondary nodes in $B_{R_I^{(sp)}}$ of a primary receiver will detect the corresponding primary transmitter and refrain from transmission. However, as shown in Section \ref{subsec:SUspatialThroughput}, increasing $R_D$ deteriorates the secondary network throughput and choosing $R_D = \w{R_r^{(p)}}$ diminishes the secondary network throughput to zero asymptotically (c.f. Lemma \ref{lemma:RDtooBig}). Therefore, in what follows, we assume $R_D = \OOO{R_r^{(p)}}$ and $R_D \leq R_I^{(sp)} + R_r^{(p)}$.
  \item [ii)] For a given $R_D \leq R_I^{(sp)} + R_r^{(p)}$, the closer $X^{(p)}_n$ is to $X^{(p)}_{n+1}$, the lower is the likelihood of interference from secondary nodes to $X^{(p)}_{n+1}$. Hence, in the overlaid scenario, $\Lambda_{X^{(p)}_n}$ and $Y_{X^{(p)}_n}$ are no longer independent and the separation principle does not directly apply.
\end{itemize}

In the following, we derive the asymptotic spatial throughput of the primary network in the presence of a secondary tier. We first consider the $\beta>1$ scenario. In this case we have $R_r^{(p)} = \w{R_r^{(s)}}$. In Propositions \ref{prop:RdTooSmall} and \ref{prop:RdBig} given below, we establish that regardless of the secondary spectrum sensing settings (i.e., $R_D = \ooo{R_r^{(p)}}$ or $R_D = \OOO{R_r^{(p)}}$), the primary network can still achieve its stand-alone sum spatial throughput scaling when $\beta>1$. Furthermore, we derive the primary network AMG and identify its relation with secondary medium access and spectrum sensing strategies.

\begin{prop}
\label{prop:RdTooSmall}
Assuming $\beta>1$ and $R_D = \ooo{R_r^{(p)}}$, the primary network throughput is asymptotically independent of the secondary network spectrum sensing and can be obtained as
\begin{align}
\label{eq:PUthroughbeta>1Rd0}
C^{(p)}_{\beta>1} \sim  \chi^{(p)}_{\{\beta > 1\}} \sqrt{\frac{\lambda^{(p)}}{\log(\lambda^{(p)})}}\,,
\end{align}
where the primary network AMG in the presence of a secondary network equals
\begin{equation}
\label{eq:PU_AMG_beta>1_RdTooSmall}
\chi^{(p)}_{\{\beta > 1\}} = \gamma^{\alpha_1} \chi\,,
\end{equation}
when the secondary medium access probability equals $q^{(s)} = \alpha_1(\lambda^{(s)} \pi (R_I^{(s)})^2)^{-1}$, with $\alpha_1 > 0$, and $\gamma := \exp(-(R_I^{(sp)}/R_I^{(s)})^2) < 1$.
\end{prop}
\begin{proof}
Let $\sigma_n$ denote the event that no primary transmitters fall into $B_{R_D+R_I^{(sp)}}(X_{n+1}^{(p)})$ and let $\hat{\Lambda}_{2,X^{(p)}_{n+1}}$ denote the event that no primary users in $B_{R_I^{(p)}}(X_{n+1}^{(p)})-B_{R_D+R_I^{(sp)}}(X_{n+1}^{(p)})$, except $X_n^{(p)}$, initiate transmissions. We have that $\Lambda_{1,X^{(p)}_n}$ and $\hat{\Lambda}_{2,X^{(p)}_{n+1}}$ are independent of $\sigma_n$, $X^{(p)}_n$, and $X^{(p)}_{n+1}$. In addition, given $\sigma_n$, the secondary users located inside $B_{R_I^{(sp)}}(X_{n+1}^{(p)})$ detect \emph{no} primary transmitters and initiate transmissions with probability $q^{(s)}$ independent of $\Lambda_{1,X^{(p)}_n}$ and $\hat{\Lambda}_{2,X^{(p)}_{n+1}}$. Together with \eqref{eq:betterdef} and \eqref{eq:1tierthroughput_0}, we have\footnote{By an abuse of notation, we abbreviate $\nu_{R_r^{(p)}}$ and $\nu_{R_r^{(s)}}$ with $\nu$ when the correct form is clear from context.}
\begin{align*}
C^{(p)}_{\beta>1} &= \lambda^{(p)}|A|\ee{\frac{1}{\nu+1}\sum_{n=0}^{\nu}\ee{Y_{X_n^{(p)}}\I{\Lambda_{X^{(p)}_n}}}} \\
&\sim \lambda^{(p)}|A|\ee{\frac{1}{\nu+1}\sum_{n=0}^{\nu}\ee[\sigma_n]{Y_{X_n^{(p)}}\I{\Lambda_{X^{(p)}_n}}}} \\
&=\lambda^{(p)}|A|\ee{\frac{1}{\nu+1}\sum_{n=0}^{\nu}\ee[\sigma_n]{Y_{X_n^{(p)}}\I{\Lambda_{1,X^{(p)}_n}\hat{\Lambda}_{2,X^{(p)}_{n+1}}}}\p{\Lambda_{3,X^{(p)}_{n+1}}}} \\
&=\lambda^{(p)}|A|\ee{\frac{1}{\nu+1}\sum_{n=0}^{\nu}\ee[\{|X_n^{(p)}-X_{n+1}^{(p)}|>R_D+R_I^{(sp)}\}]{Y_{X_n^{(p)}}}}\p{\Lambda_{1,X^{(p)}_n}\hat{\Lambda}_{2,X^{(p)}_{n+1}}}\p{\Lambda_{3,X^{(p)}_{n+1}}}\\
&\sim  C^{(p)} e^{-\lambda^{(s)} q^{(s)} (R_I^{(sp)})^2}\,,
\end{align*}
where the second line is due to $\pp{\overline{\sigma}_n} = \exp(-\lambda^{(p)} q^{(p)} (R_D+R_I^{(sp)})^2) \to 0$ and the last line is due to
\begin{align*}
\ee{Y_{X_n^{(p)}}\Big| |X_n^{(p)}-X_{n+1}^{(p)}|>R_D+R_I^{(sp)}}&\to\ee{Y_{X_n^{(p)}}}\,, \\
\pp{\hat{\Lambda}_{2,X^{(p)}_{n+1}}}&\to\pp{\Lambda_{2,X^{(p)}_{n+1}}}\,,
\end{align*}
as $\lambda^{(p)} \to \infty$ since $R_D = \ooo{R_r^{(p)}}$ and $R_r^{(sp)} = \ooo{R_r^{(p)}}$. Now choosing $q^{(s)} = \alpha_1(\lambda^{(s)} \pi (R_I^{(s)})^2)^{-1}$ with $\alpha_1>0$ and taking $\lambda^{(p)}\to\infty$, we have \eqref{eq:PU_AMG_beta>1_RdTooSmall}.
\end{proof}

From Proposition \ref{prop:RdTooSmall}, we observe that choosing $R_D = \ooo{R_r^{(p)}}$ is counter-productive and spectrum sensing cannot improve the primary network throughput. Next, we consider the case where $R_D = \OOO{R_r^{(p)}}$, and in particular assume that $R_D = \alpha_2 R_r^{(p)}$, for $0<\alpha_2\leq 1$.

\begin{prop}
\label{prop:RdBig}
Assume $\beta>1$ and $R_D = \alpha_2 R_r^{(p)}$ with $0<\alpha_2\leq 1$. Then, the primary network spatial throughput can be obtained as
\begin{align}
\label{eq:PUthroughbeta>1RdBig}
C^{(p)}_{\beta>1} \sim  \chi^{(p)}_{\{\beta > 1\}} \sqrt{\frac{\lambda^{(p)}}{\log(\lambda^{(p)})}}\,,
\end{align}
where the primary network AMG in the presence of a secondary network equals
\begin{equation}
\label{eq:PU_AMG_beta>1_RdBig}
\chi^{(p)}_{\{\beta > 1\}} = \left(\alpha_2^3 + (1-\alpha_2^3)\gamma^{\alpha_1}\right)\chi\,,
\end{equation}
when the secondary medium access probability equals $q^{(s)} = \alpha_1(\lambda^{(s)} \pi (R_I^{(s)})^2)^{-1}$, with $\alpha_1 > 0$, and $\gamma := \exp(-(R_I^{(sp)}/R_I^{(s)})^2) < 1$.
\end{prop}
\begin{proof}
Define $\sigma_{1,n} := \{|X_n^{(p)}-X_{n+1}^{(p)}| \leq R_D-R_I^{(sp)}\}$, $\sigma_{2,n} := \{R_D-R_I^{(sp)}\leq |X_n^{(p)}-X_{n+1}^{(p)}| \leq R_D+R_I^{(sp)}\}$, and $\sigma_{3,n} := \{|X_n^{(p)}-X_{n+1}^{(p)}| \geq R_D+R_I^{(sp)}\}$. Given $\sigma_{1,n}$, all secondary users in $B_{R_I^{(sp)}}(X^{(p)}_{n+1})$ will detect the transmission of $X_n^{(p)}$ and refrain from transmission. In this case, $X_{n+1}^{(p)}$ does not perceive any inter-network interference from the secondary network and we can apply the separation principle to compute the conditional spatial throughput for the primary network. Given $\sigma_{3,n}$, we have that $X_n^{(p)}$ is out of the detection ranges of all secondary users in $B_{R_I^{(sp)}}(X^{(p)}_{n+1})$ and consequently, the event $\Lambda_{X^{(p)}_n}$ is independent of $X_n^{(p)}$ and $X_{n+1}^{(p)}$. Also note that in this case, given $\Lambda_{2,X^{(p)}_{n+1}}$, the secondary users in $B_{R_I^{(sp)}}(X^{(p)}_{n+1})$ detect no primary transmitters (since $R_D+R_I^{(sp)}\leq |X_n^{(p)}-X_{n+1}^{(p)}| \leq R_r^{(p)} \leq R_I^{(p)}$) and initiate transmissions with probability $q^{(s)}$. Hence, using \eqref{eq:betterdef} we obtain the primary network spatial throughput as
\begin{align*}
C^{(p)}_{\beta>1} &= \lambda^{(p)}|A|\ee{\frac{1}{\nu+1}\sum_{n=0}^{\nu}\sum_{i=1}^3 \ee[\sigma_{i,n}]{Y_{X_n^{(p)}}\I{\Lambda_{X^{(p)}_n}}}\p{\sigma_{i,n}}} \\
&= \lambda^{(p)}|A|\mathrm{E}\Bigg(\frac{1}{\nu+1}\sum_{n=0}^{\nu} \bigg[\ee[\sigma_{1,n}]{Y_{X_n^{(p)}}}\p{\Lambda_{1,X^{(p)}_n}\Lambda_{2,X^{(p)}_{n+1}}}\p{\sigma_{1,n}} \\
&\qquad\qquad\quad+\ee[\sigma_{2,n}]{Y_{X_n^{(p)}}\I{\Lambda_{X^{(p)}_n}}}\p{\sigma_{2,n}} \\
&\qquad\qquad\quad+\ee[\sigma_{3,n}]{Y_{X_n^{(p)}}}\p{\Lambda_{1,X^{(p)}_n}\Lambda_{2,X^{(p)}_{n+1}}}\p{\Lambda_{3,X^{(p)}_{n+1}}\mid \Lambda_{2,X^{(p)}_{n+1}}}\p{\sigma_{3,n}}\bigg]\Bigg)\\
&\sim C^{(p)} \left(\alpha_2^3 + (1-\alpha_2^3) e^{-\lambda^{(s)} q^{(s)} \pi (R_I^{(sp)})^2}\right)\,,
\end{align*}
where the last line is due to
\begin{align*}
\ee[\sigma_{1,n}]{Y_{X_n^{(p)}}} &= \frac{4}{3\pi}(R_D-R_I^{(sp)})\,, \\
\ee[\sigma_{3,n}]{Y_{X_n^{(p)}}} &= \frac{4}{3\pi}\frac{(R_r^{(p)})^3-(R_D+R_I^{(sp)})^3}{(R_r^{(p)})^2-(R_D+R_I^{(sp)})^2}\,, \\
\p{\sigma_{2,n}} &= \OO{(\lambda^{(p)})^{1-\beta}}\,.
\end{align*}
\end{proof}

\begin{remark}
Observe that the spatial throughput of the primary network is strictly degraded by a constant factor asymptotically in the presence of the secondary network, i.e., $C^{(p)}_{\{\beta > 1\}} =  \frac{\chi^{(p)}_{\{\beta > 1\}}}{\chi}\,C^{(p)}$, with $0<\chi^{(p)}_{\{\beta > 1\}}/\chi \leq 1$. However, based on \eqref{eq:PU_AMG_beta>1_RdBig}, the primary network AMG loss can be recovered by decreasing $\alpha_1$ or increasing $\alpha_2$. In other words, in order to satisfy the QoS requirement (i.e., the minimum relative AMG) of the primary network, the secondary network needs to decrease its medium access probability (through decreasing $\alpha_1$) or increase its detection range (by increasing $\alpha_2$).
\end{remark}

Now, we consider primary network spatial throughput when $\beta < 1$, where we have much fewer secondary nodes with much larger interference ranges (than primary nodes) and $R_r^{(p)} = \ooo{R_r^{(s)}}$.

\begin{prop}
\label{prop:PUtrhoughput_beta<1}
Assuming $\beta<1$, the primary network spatial throughput can be obtained as
\begin{align}
\label{eq:PUthroughbeta<1}
C^{(p)}_{\beta<1} \sim  \chi^{(p)}_{\{\beta < 1\}} \sqrt{\frac{\lambda^{(p)}}{\log(\lambda^{(p)})}}\,,
\end{align}
where the primary network AMG in the presence of secondary network only depends on the effective medium access probability $\tilde{q}^{(s)}:= q^{(s)}\exp(-\lambda^{(p)}q^{(p)}\pi R_D^2)$ of secondary users:
\begin{equation}
\label{eq:PU_AMG_beta<1}
\chi^{(p)}_{\{\beta < 1\}} = e^{-\lambda^{(s)}\tilde{q}^{(s)}\pi(R_I^{(sp)})^2}\chi\,.
\end{equation}
\end{prop}
\begin{proof}
In Proposition \ref{prop:SUthroughput_beta<1}, we show that the secondary users initiate transmission with probability no less than $\tilde{q}^{(s)}$ and no greater than $\check{q}^{(s)}$ (as defined in \eqref{eq:ineqb}) when $\beta < 1$. Together with the fact that $\check{q}^{(s)}\to\tilde{q}^{(s)}$ as $\lambda^{(p)}\to\infty$, we obtain that
\begin{align*}
C^{(p)}_{\beta<1} &= \lambda^{(p)}|A|\ee{\frac{1}{\nu+1}\sum_{n=0}^{\nu}\ee{Y_{X_n^{(p)}}\I{\Lambda_{X^{(p)}_n}}}} \nonumber\\
&\sim \lambda^{(p)}|A|\ee{\frac{1}{\nu+1}\sum_{n=0}^{\nu}\ee{Y_{X_n^{(p)}}\I{\Lambda_{1,X^{(p)}_n}\Lambda_{2,X^{(p)}_{n+1}}}}}\p{\Lambda_{3,X^{(p)}_{n+1}}}\nonumber\\
&\sim C^{(p)}e^{-\lambda^{(s)}\tilde{q}^{(s)}\pi(R_I^{(sp)})^2}\,.
\end{align*}
\end{proof}

\subsection{Throughput Analysis for the Secondary Network}
\label{subsec:SUspatialThroughput}

In this section we derive the spatial throughput for the secondary network when secondary users try to access the channel opportunistically in the presence of primary users. The throughput analysis closely follows the methods in Section \ref{sec:singletier} but with proper modifications to the calculation of successful transmission probability, which now should take into account the opportunistic access mechanism adopted by secondary users and the extra inter-network interference from primary users.

Let $\tilde{\Lambda}_{X^{(s)}_n}$ be the event of successful transmission of a packet $b$ from a secondary node $X^{(s)}_n$ to the next relay $X^{(s)}_{n+1}$ in the presence of the primary network. Similar to Section \ref{subsec:PUspatialThroughput}, we have that $\tilde{\Lambda}_{X^{(s)}_n}$ happens if events $\tilde{\Lambda}_{1,X^{(s)}_n}$, $\tilde{\Lambda}_{2,X^{(s)}_{n+1}}$, and $\Lambda_{3,X^{(s)}_{n+1}}$ all happen. Here, $\Lambda_{3,X^{(s)}_{n+1}}$ denotes the event that there are no primary transmitters within inter-network interference range $R_I^{(ps)}$ of $X^{(s)}_{n+1}$. $\tilde{\Lambda}_{1,X^{(s)}_n}$ and $\tilde{\Lambda}_{2,X^{(s)}_{n+1}}$ are similar to the events in the proof of Lemma \ref{lemma:separation_principle}, except that unlike the single-tier network case, the secondary users initiate transmissions with probability $q^{(s)}$ only when they detect no primary transmitters within $R_D$ radius.

We define the effective access probability $\tilde{q}^{(s)} := q^{(s)}\exp(-\lambda^{(p)}q^{(p)}\pi R_D^2)$ and denote by $\Lambda_{X^{(s)}_n}:=\Lambda_{1,X^{(s)}_n}\Lambda_{2,X^{(s)}_{n+1}}\Lambda_{3,X^{(s)}_{n+1}}$ the event of successful transmission if all secondary users initiate transmissions with probability $q^{(s)}$ regardless of the spectrum sensing outcome. The distinctive feature in the secondary network is that the transmission initiation is contingent upon the detection of an idle spectrum. Thus, the larger $R_D$ is, the smaller the likelihood of secondary transmission initiation is. On the other hand, the larger $R_D$ is, the smaller the likelihood of intra-network interference among secondary users is. Therefore, there exists a tradeoff between $\pp{\tilde{\Lambda}_{1,X^{(s)}_n}}$ and $\pp{\tilde{\Lambda}_{2,X^{(s)}_{n+1}}}$ via the choice of $R_D$.

We show in Lemma \ref{lemma:RDtooBig} that the secondary network sum throughput is asymptotically zero when $R_D = \w{R_r^{(p)}}$ regardless of the relative density of two networks.
\begin{lemma}
\label{lemma:RDtooBig}
The secondary network sum throughput is asymptotically zero when $R_D = \w{R_r^{(p)}}$. Therefore, in order to satisfy the primary network QoS requirement (i.e., the minimum relative AMG) while achieving asymptotically non-trivial sum throughput for the secondary network, the detection range should be chosen as $R_D = \alpha_2 R_r^{(p)}$, with constant $0<\alpha_2\leq 1$ when $\beta > 1$ and $R_D = \alpha_2 R_r^{(p)}$, with constant $\alpha_2>0$ when $\beta<1$.
\end{lemma}
\begin{proof}
Using \eqref{eq:betterdef} we have
\begin{align*}
C^{(s)}_{\beta>1} &= \lambda^{(s)}|A|\ee{\frac{1}{\nu+1}\sum_{n=0}^{\nu}\ee{Y_{X_n^{(s)}}\I{\tilde{\Lambda}_{X^{(s)}_n}}}} \\
& \leq \lambda^{(s)}|A| R_r^{(s)} \p{\tilde{\Lambda}_{1,X^{(s)}_n}} \\
%& \leq \lambda^{(s)}|A| R_r^{(s)} \p{\tilde{\Lambda}_{X^{(s)}_n}\mid\tilde{\Lambda}_{1,X^{(s)}_n}}\p{\tilde{\Lambda}_{1,X^{(s)}_n}} \\
& \leq \lambda^{(s)}|A| R_r^{(s)} e^{-\big(\frac{R_D}{R_I^{(p)}}\big)^2}\to 0
\end{align*}
as $\lambda^{(p)}\to\infty$. Together with the result in Proposition \ref{prop:RdTooSmall}, the proof of the lemma is complete.
\end{proof}

In the following, we derive the asymptotic spatial throughput of the secondary network in the presence of a primary tier when the secondary spectrum sensing range is set as $R_D = \alpha_2 R_r^{(p)}$, with constant $0<\alpha_2\leq 1$ when $\beta > 1$ and $R_D = \alpha_2 R_r^{(p)}$, with constant $\alpha_2>0$ when $\beta<1$. We first consider the $\beta>1$ scenario. In this case we have $R_r^{(p)} = \w{R_r^{(s)}}$. As mentioned before, the medium access decisions of secondary users are correlated due to their overlapping spectrum sensing regions. For example, if $X_n^{(s)}$ initiates a transmission, the probability that the secondary users located inside $B_{R_I^{(s)}}(X_{n+1}^{(s)})$ initiate transmissions increases, which in turn decreases the probability of successful transmissions between $X_n^{(s)}$ and $X_{n+1}^{(s)}$. Furthermore, as $X_n^{(s)}$ gets closer to $X_{n+1}^{(s)}$, the probability of intra-network interference to $X_{n+1}^{(s)}$ increases, knowing $X_n^{(s)}$ initiates transmission.

In general, the probability that a secondary node $X^{(s)}_i$ initiates a transmission is a non-increasing function of $|X^{(s)}_j-X^{(s)}_i|$ if $X^{(s)}_j$ is transmitting and a non-decreasing function of $|X^{(s)}_j-X^{(s)}_i|$ if $X^{(s)}_j$ is idling. Similarly, the probability that a secondary node $X^{(s)}_i$ idles is a non-decreasing function of $|X^{(s)}_j-X^{(s)}_i|$ if $X^{(s)}_j$ is transmitting and a non-increasing function of $|X^{(s)}_j-X^{(s)}_i|$ if $X^{(s)}_j$ is idling.

In Propositions \ref{prop:SUthroughput_beta>1}, we establish that the secondary network can still achieve its stand-alone sum spatial throughput scaling when $\beta>1$. Furthermore, we derive the secondary network AMG and identify its relation with the secondary medium access and spectrum sensing strategies.

\begin{prop}
\label{prop:SUthroughput_beta>1}
Assume $\beta>1$. The secondary network sum spatial throughput can be obtained as
\begin{align}
\label{eq:SUthroughbeta>1}
C^{(s)}_{\beta>1} \sim  \chi^{(s)}_{\{\beta > 1\}} \sqrt{\frac{\lambda^{(s)}}{\log(\lambda^{(s)})}}\,,
\end{align}
where the secondary network AMG in the presence of primary network equals
\begin{equation}
\label{eq:SU_AMG_beta>1}
\chi^{(s)}_{\{\beta > 1\}} = e^{-\big(\frac{\max\{\alpha_2 R_r^{(p)},R_I^{(ps)}\}}{R_I^{(p)}}\big)^2}\alpha_1e^{1-\alpha_1}\chi\,,
\end{equation}
when the secondary medium access probability equals $q^{(s)} = \alpha_1(\lambda^{(s)} \pi (R_I^{(s)})^2)^{-1}$ and $R_D = \alpha_2 R_r^{(p)}$, with $\alpha_1>0$ and $0<\alpha_2\leq 1$.
\end{prop}
\begin{proof}
Let us first consider the case where $R_D \leq R_I^{(ps)}-R_I^{(s)}$. In this case, given $\Lambda_{3,X^{(s)}_{n+1}}$, all secondary users in $B_{R_I^{(s)}}(X_{n+1}^{(s)})$ together with $X_n^{(s)}$ and $X_{n+1}^{(s)}$ detect no primary users and independently initiate transmissions with probability $q^{(s)}$. Hence, using the separation principle and \eqref{eq:betterdef}, we have
\begin{align}
\label{eq:SU_AMG_beta>1_a}
C^{(s)}_{\beta>1} &= \lambda^{(s)}|A|\ee{\frac{1}{\nu+1}\sum_{n=0}^{\nu}\ee{Y_{X_n^{(s)}}\I{\tilde{\Lambda}_{X^{(s)}_n}}}} \nonumber\\
&= \lambda^{(s)}|A|\ee{\frac{1}{\nu+1}\sum_{n=0}^{\nu}\ee[\Lambda_{3,X^{(s)}_{n+1}}]{Y_{X_n^{(s)}}\I{\tilde{\Lambda}_{1,X^{(s)}_n}\tilde{\Lambda}_{2,X^{(s)}_{n+1}}}}}\p{\Lambda_{3,X^{(s)}_{n+1}}} \nonumber\\
&= \lambda^{(s)}|A|\ee{\frac{1}{\nu+1}\sum_{n=0}^{\nu}\ee[\Lambda_{3,X^{(s)}_{n+1}}]{Y_{X_n^{(s)}}\I{\Lambda_{1,X^{(s)}_n}\Lambda_{2,X^{(s)}_{n+1}}}}}\p{\Lambda_{3,X^{(s)}_{n+1}}} \nonumber\\
&= C^{(s)} e^{-(\frac{R_I^{(ps)}}{R_I^{(p)}})^2}\,.
\end{align}

Now consider the case with $R_D > R_I^{(ps)}-R_I^{(s)}$. In this case, observe that $\tilde{\Lambda}_{1,X^{(s)}_n}\tilde{\Lambda}_{2,X^{(s)}_{n+1}}\hat{\Lambda}_{3,X^{(s)}_{n+1}} \subseteq \tilde{\Lambda}_{1,X^{(s)}_n}\tilde{\Lambda}_{2,X^{(s)}_{n+1}}\Lambda_{3,X^{(s)}_{n+1}}$, where $\hat{\Lambda}_{3,X^{(s)}_{n+1}}$ denotes the event that there are no primary transmitters within a $R_D + R_I^{(s)}$ radius of $X^{(s)}_{n+1}$; again, given $\hat{\Lambda}_{3,X^{(s)}_{n+1}}$, $\tilde{\Lambda}_{1,X^{(s)}_n}\tilde{\Lambda}_{2,X^{(s)}_{n+1}}$  is independent of the spectrum sensing outcome. Hence, using the separation principle, we obtain the following lower bound for the secondary network sum spatial throughput:
\begin{align}
\label{eq:SUthroughLowerBound_RDbig}
C^{(s)}_{\beta>1} &= \lambda^{(s)}|A|\ee{\frac{1}{\nu+1}\sum_{n=0}^{\nu}\ee{Y_{X_n^{(s)}}\I{\tilde{\Lambda}_{X^{(s)}_n}}}} \nonumber\\
&\geq \lambda^{(s)}|A|\ee{\frac{1}{\nu+1}\sum_{n=0}^{\nu}\ee[\hat{\Lambda}_{3,X^{(s)}_{n+1}}]{Y_{X_n^{(s)}}\I{\tilde{\Lambda}_{1,X^{(s)}_n}\tilde{\Lambda}_{2,X^{(s)}_{n+1}}}}\p{\hat{\Lambda}_{3,X^{(s)}_{n+1}}}} \nonumber\\
&=\lambda^{(s)}|A|\ee{\frac{1}{\nu+1}\sum_{n=0}^{\nu}\ee{Y_{X_n^{(s)}}\I{\Lambda_{1,X^{(s)}_n}\Lambda_{2,X^{(s)}_{n+1}}}}\p{\hat{\Lambda}_{3,X^{(s)}_{n+1}}}} \nonumber\\
&=  C^{(s)} e^{-(\frac{R_D}{R_I^{(p)}})^2}\,.
\end{align}

Next we derive an upper bound for the secondary network sum spatial throughput. Assume there are $N$ secondary users $\{X_1,X_2,\ldots, X_N\}$ located inside $B_{R_I^{(s)}}(X_{n+1}^{(s)})$ including $X_{n+1}^{(s)}$ itself and excluding $X_n^{(s)}$. Let $\sigma_i$ denote the event that $X_i$ initiates a transmission, $\overline{\sigma}_i$ denotes the event that $X_i$ remains silent, and $\sigma_{-i}$ denote the event that at least one of $\{X_1,X_2,\ldots, X_N\} \backslash \{X_i\}$ initiate a transmission. Given $\Lambda_{3,X^{(s)}_{n+1}}$ and $\tilde{\Lambda}_{1,X^{(s)}_n}$, the probability that $X_i$ remains idle is no more than $\pp{\overline{\sigma}_i} \leq 1-\hat{q}^{(s)}$ (i.e., when $X_i$ and $X^{(s)}_n$ are farthest away) and no less than $\pp{\overline{\sigma}_i} \geq 1-q^{(s)}$, where $\hat{q}^{(s)} := q^{(s)}\exp(-\lambda^{(p)}q^{(p)}|B_{R_D}(R_I^{(s)},0)-B_{R_D}(-R_I^{(s)},0)-B_{R_I^{(ps)}}(0,0)|)$. Similarly, given $\Lambda_{3,X^{(s)}_{n+1}}$ and $\tilde{\Lambda}_{1,X^{(s)}_n}$, the probability that $X_i$ initiates a transmission is no less than $\pp{\sigma_i}\geq \hat{q}^{(s)}$. Consequently, given $\Lambda_{3,X^{(s)}_{n+1}}$ and $\tilde{\Lambda}_{1,X^{(s)}_n}$, we have
\begin{align}
\label{eq:NoSUtrans_prim}
N (1-\hat{q}^{(s)}) \geq \sum_{i=1}^{N} \p{\overline{\sigma}_i} &= \sum_{i=1}^{N} \pp{\overline{\sigma}_i\cap \overline{\sigma}_{-i}} +  \sum_{i=1}^{N} \p{\overline{\sigma}_i\cap\sigma_{-i}} \nonumber\\
&= \sum_{i=1}^{N}\pp{\bigcap_{j=1}^N \overline{\sigma}_j} + \sum_{i=1}^{N}\p{\overline{\sigma}_i}\p{\sigma_{-i}\mid\overline{\sigma}_i} \nonumber\\
&\geq N \pp{\bigcap_{j=1}^N \overline{\sigma}_j} + N(1-q^{(s)}) \sum_{j=1}^{N-1} {N-1 \choose j} (\hat{q}^{(s)})^{j} (1-q^{(s)})^{N-1-j} \nonumber\\
&= N \pp{\bigcap_{j=1}^N \overline{\sigma}_j} + N(1-q^{(s)}) \big[(1-q^{(s)} + \hat{q}^{(s)})^{N-1} - (1-q^{(s)})^{N-1}\big]\,.
\end{align}

Taking expectation over the number of nodes falling inside $B_{R_I^{(s)}}(X_{n+1}^{(s)})$, we obtain
\begin{align}
\label{eq:NoSUtrans}
\p{\tilde{\Lambda}_{2,X^{(s)}_{n+1}}\mid\tilde{\Lambda}_{1,X^{(s)}_n}\Lambda_{3,X^{(s)}_{n+1}}} &= \eee{\pp{\bigcap_{i=1}^N \overline{\sigma}_i\mid\Lambda_{3,X^{(s)}_{n+1}}\tilde{\Lambda}_{1,X^{(s)}_n}}} \nonumber\\
&\leq e^{-\lambda^{(s)}q^{(s)}\pi(R_I^{(s)})^2} + \left[(1-\hat{q}^{(s)}) - (1-q^{(s)}) e^{-\lambda^{(s)}(q^{(s)}-\hat{q}^{(s)})\pi(R_I^{(s)})^2}\right]\,.
\end{align}

Hence, we have
\begin{align}
\label{eq:SUthroughUpperBound_RDbig}
C^{(s)}_{\beta>1} &= \lambda^{(s)}|A|\ee{\frac{1}{\nu+1}\sum_{n=0}^{\nu}\ee{Y_{X_n^{(s)}}\I{\tilde{\Lambda}_{X^{(s)}_n}}}} \nonumber\\
&\leq \lambda^{(s)}|A| \ee{\frac{1}{\nu+1}\sum_{n=0}^{\nu}\ee[\tilde{\Lambda}_{1,X^{(s)}_n}\Lambda_{3,X^{(s)}_{n+1}}X^{(s)}_nX^{(s)}_{n+1}]{Y_{X_n^{(s)}}\I{\tilde{\Lambda}_{2,X^{(s)}_{n+1}} }}q^{(s)}e^{-(\frac{R_D}{R_I^{(p)}})^2}} \nonumber\\
&\leq \lambda^{(s)}|A|q^{(s)}e^{-(\frac{R_D}{R_I^{(p)}})^2} \ee{\frac{1}{\nu+1}\sum_{n=0}^{\nu}\ee{Y_{X_n^{(s)}}}}\max\limits_{X^{(s)}_n,X^{(s)}_{n+1}} \p{\tilde{\Lambda}_{2,X^{(s)}_{n+1}}\mid\tilde{\Lambda}_{1,X^{(s)}_n}\Lambda_{3,X^{(s)}_{n+1}}X^{(s)}_nX^{(s)}_{n+1}} \nonumber\\
&= C^{(s)} e^{-(\frac{R_D}{R_I^{(p)}})^2} \left[1+(1-\hat{q}^{(s)})e^{\lambda^{(s)}q^{(s)}\pi(R_I^{(s)})^2} - (1-q^{(s)}) e^{\lambda^{(s)}\hat{q}^{(s)}\pi(R_I^{(s)})^2}\right]\,.
\end{align}

From \eqref{eq:SUthroughLowerBound_RDbig}, \eqref{eq:SUthroughUpperBound_RDbig}, and the fact that $\hat{q}^{(s)}\to q^{(s)}$ as $\lambda^{(p)}\to\infty$, we conclude that $C^{(s)}_{\beta>1} \sim C^{(s)} \exp(-(\frac{R_D}{R_I^{(p)}})^2)$ when $R_D > R_I^{(ps)}-R_I^{(s)}$. Finally, together with \eqref{eq:SU_AMG_beta>1_a}, we obtain \eqref{eq:SU_AMG_beta>1}.
\end{proof}

\begin{remark}
From \eqref{eq:SU_AMG_beta>1} we have that $q^{(s)} = \OO{1/\log(\lambda^{(s)})}$ is still a necessary condition to ensure an asymptotically nontrivial throughput for the secondary network. Similar to the single-tier network case, setting $q^{(s)} = (\lambda^{(s)} \pi (R_I^{(s)})^2)^{-1}$ is still the optimal access probability for the secondary nodes. Observe that setting $\alpha_1 \neq 1$ or $\alpha_2 > R_I^{(ps)}/R_r^{(p)}$ degrades the secondary network AMG. However, the secondary network AMG remains unaffected for $\alpha_2 < R_I^{(ps)}/R_r^{(p)}$.  Furthermore, recall from \eqref{eq:PU_AMG_beta>1_RdBig}, the primary network AMG can be recovered by increasing $\alpha_1>1$ or $\alpha_2$.
\end{remark}

Next, we determine the secondary network throughput scaling and AMG when $\beta < 1$. In this case we have $R_r^{(p)} = \ooo{R_r^{(s)}}$. In the next proposition, we derive the secondary network spatial throughput and show that the secondary network can still achieve its stand-alone sum spatial throughput scaling when $\beta<1$.
\begin{prop}
\label{prop:SUthroughput_beta<1}
When $\beta<1$, the secondary network throughput performance in the presence of primary users resembles the stand-alone secondary network but with a reduced medium access probability $\tilde{q}^{(s)}$. In other words,
\begin{equation}
\label{eq:SU_AMG_beta<1}
C^{(s)}_{\{\beta<1\}} =  e^{-(\frac{R_I^{(ps)}}{R_I^{(p)}})^2} \tilde{C}^{(s)}\,,
\end{equation}
where $\tilde{C}^{(s)}$ equals the single-tier spatial throughput expression in \eqref{eq:1tierthroughput_0} with the secondary network parameters and the effective medium access probability $\tilde{q}^{(s)}$ substituted. The secondary network can achieve a throughput scaling of $\T{\sqrt{\lambda^{(s)}/\log(\lambda^{(s)})}}$ with the effective ALOHA access probability $\tilde{q}^{(s)}= \OO{1/\log(\lambda^{(s)})}$ even when the secondary nodes are much more sparsely distributed than the primary nodes.
\end{prop}
\begin{proof}
Similar to the proof of Proposition \ref{prop:SUthroughput_beta>1}, assume there are $N$ secondary users $\{X_1,X_2,\ldots, X_N\}$ located inside $B_{R_I^{(s)}}(X_{n+1}^{(s)})$ including $X_{n+1}^{(s)}$ itself and excluding $X_n^{(s)}$. Define $\sigma_i$, $\overline{\sigma}_i$, and $\sigma_{-i}$ as before. Let $\varsigma_i$ denote the event that there are no \emph{secondary} users located inside $B_{2R_D}(X_i)$. Similar to \eqref{eq:NoSUtrans_prim}, \eqref{eq:NoSUtrans}, and using the facts that
\begin{subequations}
\begin{align}
\p{\sigma_i} &\geq \tilde{q}^{(s)}\,, \label{eq:ineqa}\\
\p{\sigma_i} &= \p{\sigma_i\mid\varsigma_i}\p{\varsigma_i} + \p{\sigma_i\mid\overline{\varsigma_i}}\p{\overline{\varsigma_i}} \nonumber\\
&\leq q^{(s)}\left(1-e^{-\lambda^{(s)}\pi(2R_D)^2}\right) + \tilde{q}^{(s)}e^{-\lambda^{(s)}\pi(2R_D)^2} =: \check{q}^{(s)}\,, \label{eq:ineqb}
\end{align}
\end{subequations}
when $\beta<1$, we obtain
\begin{subequations}
\begin{align}
\p{\tilde{\Lambda}_{X^{(s)}_n}} &\leq \check{q}^{(s)}\left(e^{-\lambda^{(s)}\check{q}^{(s)}\pi(R_I^{(s)})^2} + \left[(1-\tilde{q}^{(s)}) - (1-\check{q}^{(s)}) e^{-\lambda^{(s)}(\tilde{q}^{(s)}-\check{q}^{(s)})\pi(R_I^{(s)})^2}\right]\right)e^{-(\frac{R_I^{(ps)}}{R_I^{(p)}})^2}\,, \label{eq:SUsuccessbounds_beta<1_a} \\
\p{\tilde{\Lambda}_{X^{(s)}_n}} &\geq \tilde{q}^{(s)}\left(e^{-\lambda^{(s)}\tilde{q}^{(s)}\pi(R_I^{(s)})^2} + \left[(1-\check{q}^{(s)}) - (1-\tilde{q}^{(s)}) e^{-\lambda^{(s)}(\check{q}^{(s)}-\tilde{q}^{(s)})\pi(R_I^{(s)})^2}\right]\right)e^{-(\frac{R_I^{(ps)}}{R_I^{(p)}})^2}\,.\label{eq:SUsuccessbounds_beta<1_b}
\end{align}
\end{subequations}

Note that $\check{q}^{(s)}\to\tilde{q}^{(s)}$ as $\lambda^{(p)}\to\infty$ since $R_D = \OOO{R_r^{(p)}}$ and $R_r^{(p)} = \ooo{R_r^{(s)}}$ when $\beta<1$. As such, similar to \eqref{eq:SUthroughUpperBound_RDbig} we obtain
\begin{equation*}
C^{(s)}_{\beta<1} = \lambda^{(s)}|A|\ee{\frac{1}{\nu+1}\sum_{n=0}^{\nu}\ee{Y_{X_n^{(s)}}\I{\tilde{\Lambda}_{X^{(s)}_n}}}} \sim \tilde{C}^{(s)}e^{-(\frac{R_I^{(ps)}}{R_I^{(p)}})^2}\,.
\end{equation*}
\end{proof}

%\begin{remark}
%The secondary network throughput maximizes when $\tilde{q}^{(s)} = (\lambda^{(s)} \pi (R_I^{(s)})^2)^{-1}$. This requires the secondary users to access the channel with a higher probability $q^{(s)} = \frac{\exp((R_D/R_I^{(p)})^2)}{\lambda^{(s)} \pi (R_I^{(s)})^2}$ in the presence of primary users compared to the stand alone case. However, this reduces the primary network AMG by a factor of $\exp(-(R_I^{(sp)}/R_I^{(s)})^2)<e^{-1}$ (recall \eqref{eq:PU_AMG_beta<1}).
%\end{remark}
\begin{remark}
Observe that according to \eqref{eq:PU_AMG_beta<1} and \eqref{eq:SU_AMG_beta<1}, the primary and secondary network throughput performance (i.e., AMGs) depends only on $\tilde{q}^{(s)}$ when $\beta<1$. Therefore, the desired QoS requirements of the two networks can be achieved by setting either $q^{(s)}$ or $R_D$ appropriately. Hence, the spectrum sensing turns out to be unnecessary when $\beta<1$ and the desired performance can be achieved by setting the secondary network medium access parameter accordingly. In other words, it is favorable for the secondary network to blindly access the channel according to the traditional ALOHA medium access scheme without resorting to spectrum sensing when they are much sparser than the primary users.
\end{remark}

\section{Conclusion}
\label{sec:conclusion}

We studied the interaction between two overlaid ad-hoc networks: one with legacy primary users who are licensed to access the spectrum and the other with cognitive secondary users who opportunistically access the spectrum. We showed that if the secondary network is denser than the primary network, we can guarantee the same throughput scaling for both networks as that for a single network, as long as they deploy proper random access schemes. Furthermore, with the newly defined performance metric, the asymptotic multiplexing gain (AMG), we quantified how the asymptotic network performance is affected by the mutual interference between the two networks. In addition, for the first time to our knowledge, we studied the throughput performance of an overlaid cognitive network in which secondary nodes are less densely distributed than the primary users and showed that even in this scenario, both networks can achieve the single-network throughput scaling. However, unlike the case of a denser secondary network, it is possible to satisfy the QoS requirements of the licensed users without employing spectrum sensors.

%\section*{Acknowledgment}


\ifCLASSOPTIONcaptionsoff
  \newpage
\fi

\end{document}